\documentclass[lettersize,journal]{IEEEtran}
\usepackage{amsmath,amsfonts}
\usepackage{algorithmic}
\usepackage{algorithm}
\usepackage{array}
\usepackage[caption=false,font=normalsize,labelfont=sf,textfont=sf]{subfig}
\usepackage{textcomp}
\usepackage{stfloats}
\usepackage{url}
\usepackage{booktabs}
\usepackage{multirow}
\usepackage{verbatim}
\usepackage{graphicx}
\usepackage{cite}
\usepackage{balance}
\usepackage{flushend}

\begin{document}

\title{Graph Neural Network Backend for Speaker Recognition}

\author{Liang He, Ruida Li, and Mengqi Niu
  \thanks{
    Liang He, Ruida Li and Mengqi Niu are with School of Information Science and Engineering, Xinjiang University, Urumqi 830046, China.
    Liang He is also with the Department of Electronic Engineering, Tsinghua University, Beijing, China.}
  \thanks{Correspondence: heliang@mail.tsinghua.edu.cn}
}

\markboth{IEEE Signal Processing Letters,~Vol.~x, No.~x, Octorber~2022}%
	 {Shell \MakeLowercase{\textit{et al.}}: Graph Neural Network Backend for Speaker Recognition}
	 
	 \maketitle
	 
	 \begin{abstract}
    Currently, most speaker recognition backends, such as cosine, linear discriminant analysis (LDA), or probabilistic linear discriminant analysis (PLDA), make decisions by calculating similarity or distance between enrollment and test embeddings which are already extracted from neural networks. 
    However, for each embedding, the local structure of itself and its neighbor embeddings in the low-dimensional space is different, which may be helpful for the recognition but is often ignored. 
    In order to take advantage of it, we propose a graph neural network (GNN) backend to mine latent relationships among embeddings for classification. 
    We assume all the embeddings as nodes on a graph, and their edges are computed based on some similarity function, such as cosine, LDA+cosine, or LDA+PLDA. 
    We study different graph settings and explore variants of GNN to find a better message passing and aggregation way to accomplish the recognition task. 
    Experimental results on NIST SRE14 i-vector challenging, VoxCeleb1-O, VoxCeleb1-E, and VoxCeleb1-H datasets demonstrate that our proposed GNN backends significantly outperform current mainstream methods.
	 \end{abstract}
	 
	 \begin{IEEEkeywords}
	   Speaker recognition, graph neural network, embeddings, representative learning
	 \end{IEEEkeywords}
	 
	 \section{Introduction}
	 \IEEEPARstart{T}{h}e core task of speaker recognition is to determine whether two utterances are from the same speaker.
	 Currently, the mainstream methods are variants of x-vector \cite{8461375}, which has obtained excellent performance in recent evaluations and applications \cite{alenin21_interspeech}.  
	 It mainly consists of a frontend neural network responsible for mapping from an utterance with variable duration to a fixed dimension embedding, also termed an x-vector, and a backend module in charge of making the decision based on enrollment and test embeddings. 
	 
	 Most studies are about improvements of the neural networks, e.g., time-delay neural network (TDNN) \cite{8461375}, emphasized channel attention, propagation and aggregation-TDNN (ECAPA-TDNN) \cite{desplanques20_interspeech}, ResNet \cite{He2016}, ResNeXt\cite{Zhou2021}, of pooling layer, e.g., attentive statistics pooling (ASP) \cite{Safari2020}, multi-head attention pooling (MHAP) \cite{India2019}, learnable dictionary encoding (LDE) \cite{Cai2018}, and of the loss function, e.g., angular softmax (A-softmax) loss \cite{Li2018}, additive margin softmax loss (AM-softmax) \cite{Wang2018,Liu2019a}, additive angular margin (ArcFace) loss \cite{Deng2019}, dynamic margin softmax loss \cite{Zhou2020}, adaptive margin circle loss\cite{9413832}, and \emph{etc}.
	 
	 In contrast, there is less research on the backend. 
	 The mainstream backends are still cosine scoring and linear discriminant analysis (LDA) followed by probabilistic LDA (PLDA, LDA+PLDA), which is already verified on most databases or evaluations \cite{alenin21_interspeech,lozanodiez20_interspeech,shen20b_interspeech}.
	 In recent years, there are three kinds of representative methods to improve the backend.
	 The first category is about PLDA, such as Neural PLDA \cite{ramoji20_interspeech}, discriminative PLDA (DPLDA) \cite{Ferrer2020}, heavy-tailed PLDA (HT-PLDA) \cite{Silnova2018}, 
	 multi-objective optimization training of PLDA (Mot-PLDA) \cite{He2019} and \emph{etc}.
	 The second is to add an additional trainable neural network module, e.g., decision residual networks (Dr-vectors) \cite{Pelecanos2021}, deep learning backend (DLB) \cite{Ghahabi2017} and tied variational autoencoder (TVAE) \cite{Villalba2017}.
	 And the last is to develop a robust backend against domain mismatch, such as Coral++ \cite{Li2022}, domain-aware batch normalization (DABN) and domain-agnosticinstance normalization (DAIN) \cite{Hu2022}, information-maximized variational domain adversarial neural network (InfoVDANN) \cite{Tu2020}, and \emph{etc}.
	 However, these algorithms rarely use spatial or graph information among the extracted embeddings, which may significantly boost the performance.
	 
	 Recently, graph neural network (GNN) has achieved great success in a large number of areas, such as physics, chemistry, biology, knowledge graph, social network, recommendation systems, and \emph{etc} \cite{Zhou2020a}.
	 It is a powerful tool to mine rich relation information among data, which has great potential for speaker recognition. 
	 Jung \emph{et al.} propose a graph attention network (GAT) in the case of test time augmentation (TTA) \cite{Jung2021} and demonstrate that the GAT-TTA backend has consistent improvement over cosine scoring. 
	 Although the proposed GAT-TTA framework takes multiple embeddings to construct graphs, they are still only from the enrollment and test utterances, which do not use the relationship between the concerned embedding and its surrounding embeddings lying on the hypothesized hypersphere. 
	 Wang \emph{et al.} \cite{Wang2020} use a graph neural network for better clustering to accomplish the speaker diarization task. 
	 Furthermore, Zheng \emph{et al.} \cite{Lian2020} construct a heterogeneous graph to realize multi-modal information aggregation.
	 It takes the speaker and speech segment as vertexes and uses the contextual connection of the speech segment and speaker identity to calculate edges. 
	 Experimental results on the MELD databases show the effectiveness of the proposed method \cite{Lian2020}. 
  
  To take advantage of the uniqueness of each speaker's low dimensional spatial (graph) structure embedded on the hypothesized hypersphere, we propose a graph neural network (GCN) backend to mine latent relationships between embeddings and their neighbors to improve the system performance. 
	 We assume all the embeddings as nodes on a graph, and the edges are constructed by calculating the similarity between two nodes. 
	 The similarity function could be cosine, LDA+cosine, LDA+PLDA, or others, which will be examined in section \ref{sec:exp}.
	 The graph construction methods and variants of GCN are studied and compared to find a better message passing and aggregation way.
	 The experiment results on the NIST SRE14 i-vector challenging and Voxceleb-1 database validate the effectiveness of our proposed method.
	 
	 \section{Graph neural network backend for speaker recognition}
	 \subsection{Motivation}
\begin{figure}[!t]
	   \centering
	   \includegraphics[width=2.5in]{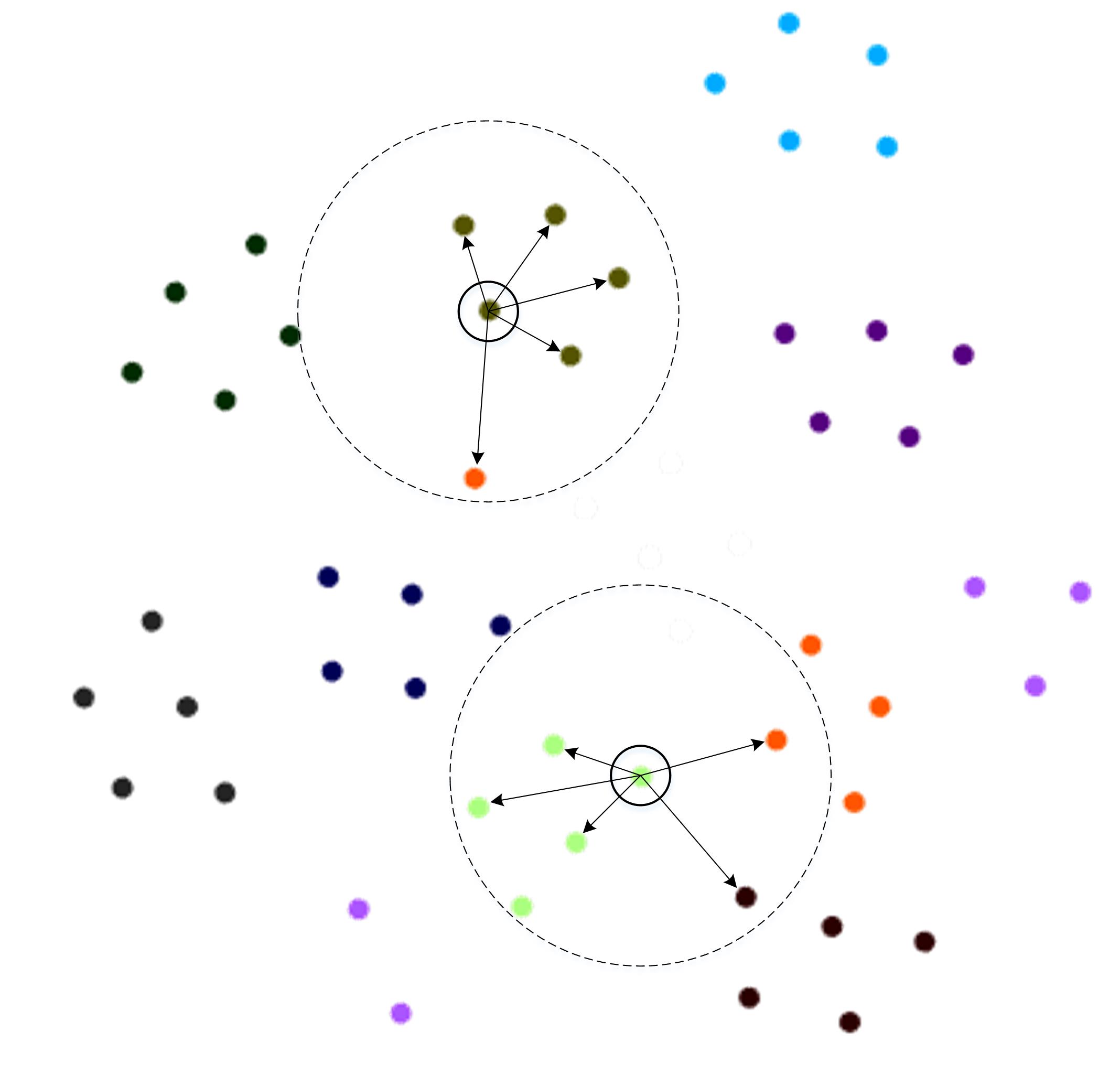}
	   \caption{We randomly select ten speakers' i-vectors from the NIST SRE14 database to visualize the relationship among them by the tSNE. We can see that both location (ellipses) and graph structure (arrows) contain discriminant information, which helps the classification.}
	   \label{fig:ebd_tsne}
	 \end{figure}

	 The task of the backend for speaker recognition is to make correct and robust decisions based on the extracted i-vectors \cite{Dehak2011} or x-vectors.
	 If we view these vectors as points in space, the position of each point and its local spatial structure with other points together will help to determine its corresponding category, see Fig. \ref{fig:ebd_tsne}. 
	 We take each point as a node and add edges by their pairwise geometric distance.
	 Thanks to the powerful ability of graph neural networks to process complex non-Euclidean data, we can more effectively use the spatial structural information on the built graph to compare them and make decisions.
	 
\begin{figure}[!t]
	   \centering
	   \includegraphics[width=3.5 in]{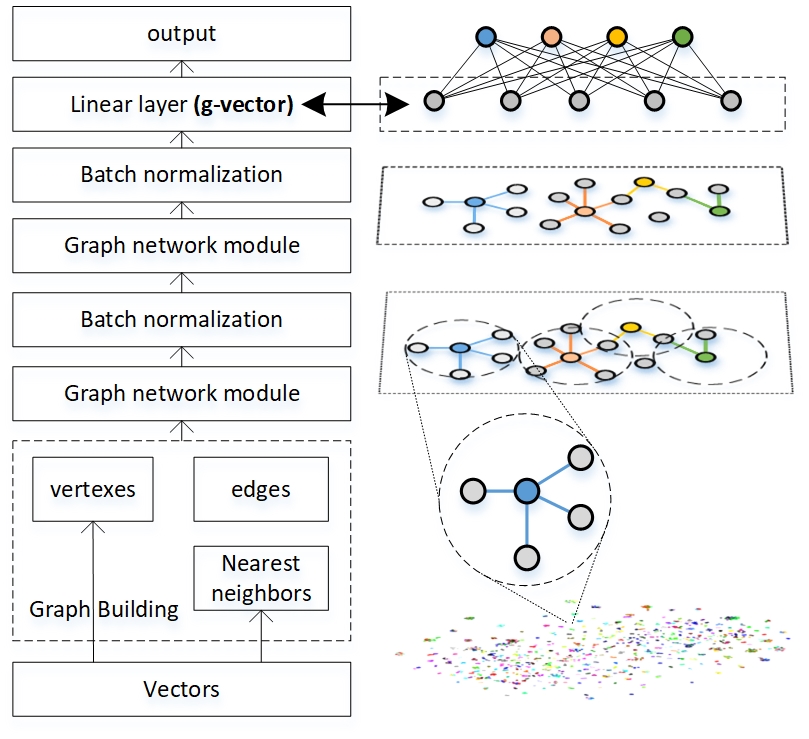}
	   \caption{
	     In the graph building, the nodes are i-vectors or x-vectors, and for each node, its corresponding edges are calculated by using the nearest neighbor algorithms. 
	     Based on the built graph, we adopt several layers of a graph network module followed by batch normalization. 
	     The last two layers include a linear layer and a softmax output layer. 
	     The number of output nodes corresponds to the speaker number in the training set.
	     In the enrollment and test phase, we read out the embeddings in the linear layer, name them as g-vectors, and use them instead of x-vectors or i-vectors for the final cosine scoring. 
	   }
	   \label{fig:gvec}
	 \end{figure}

	 Our proposed method contains graph construction and graph neural networks, see Fig. \ref{fig:gvec}. 
	 During graph construction, we mainly finish the computation of nodes and edges. 
	 The graph neural network includes graph network modules, batch normalizations (BN), fully connected layers, softmax output, and CE loss.
	 
	 \subsection{Graph building with nearest neighbors}
	 Suppose the connected graph is represented as $G=(V, E) \text{, where } V$ are the set of nodes and $E$ are the set of edges.
	 Each vector is a node on the graph, and its connected nodes are calculated by the nearest neighbor algorithm with cosine, LDA, or LDA + PLDA distance. 
	 For instance, if the cosine value of two vectors is greater than a pre-defined threshold, we add an edge to connect them.
	 After the construction of the graph, the adjacency matrix $A$ is subsequently calculated, in which the element $a_{i, j} = 1$ if there is an edge between the nodes $i$ and $j$, otherwise $a_{i, j} = 0$.
	 The diagonal elements in the adjacency matrix are set to be $1$ for including the center vectors when aggregating information. 
	 
	 \subsection{Variants of graph module}
	 Variants of graph neural network modules are different ways of message-passing to generate the next layer's nodes by aggregating nodes and their neighbors' information. 
	 Denote the $i$-th node in the $k$-th layer by $\mathbf{x}_{i}^{(k)}$, the message-passing module can be described as:
	 \begin{displaymath}
	   \mathbf{x}_{i}^{(k)}=f_{\mathbf{\Theta}}\left(\mathbf{x}_{i}^{(k-1)}, \text{AGGREGATE} (\{\mathbf{x}_{j}^{(k-1)} | j \in \mathcal{N}_{i} \}  ) \right)
	 \end{displaymath}
	 \noindent where $\mathcal{N}_i = \{j \in V| (i,j) \in E\}$ is the neighbor set of $i$-th node. $\mathbf{\Theta}$ is the parameters of message-passing module.
	 The design of $f$ and AGGREGATE is what mostly distinguishes one type of GNN from the other.
	 
	 \subsubsection{Graph convolutional network}
	 The graph convolutional network (GCN) can leverage the graph structure and aggregate node information from the neighborhoods in a convolutional way \cite{kipf2017semisupervised}. 
	 The core layer of GCN is as follows:
	 \begin{displaymath}
	   X^{(k)}=\sigma\left(\tilde{D}^{-\frac{1}{2}} \tilde{A} \tilde{D}^{-\frac{1}{2}} X^{(k-1)} \mathbf{\Theta}^{(k-1)}\right)
	 \end{displaymath} 
	 \noindent where each row of $X$
	 is $\mathbf{x}$, $\sigma$ represents the  nonlinear activation function, $\hat{A} = A + I$ denotes the adjacency matrix with inserted self-loops and $\hat{D}_{ii} = \sum_{j=0} \hat{A}_{ij}$ is its diagonal degree matrix.
	 
	 \subsubsection{Graph attention networks}
	 The graph attention networks (GAT) can leverage masked self-attentional weights to aggregate information \cite{velickovic2017graph}. 
	 The core layer of GAT is as follows:
	 \begin{displaymath}
	   \mathbf{x}^{(k)}_i = \alpha_{i,i}\mathbf{\Theta}\mathbf{x}_{i}^{(k-1)} + \sum_{j \in \mathcal{N}(i)} \alpha_{i,j}\mathbf{\Theta}\mathbf{x}_{j}^{(k-1)}   
	 \end{displaymath}
	 \noindent where the attention coefficients \(\alpha_{i,j}\) are computed as
	 \begin{displaymath}
	   \alpha_{i,j} = \frac{ \exp\left(\mathrm{LeakyReLU}\left(\mathbf{a}^{t} [\mathbf{\Theta}\mathbf{x}_i \, \| \, \mathbf{\Theta}\mathbf{x}_j] \right)\right)} {\sum_{k \in \mathcal{N}(i) \cup \{ i \}} \exp\left(\mathrm{LeakyReLU}\left(\mathbf{a}^{t} [\mathbf{\Theta}\mathbf{x}_i \, \| \, \mathbf{\Theta}\mathbf{x}_k] \right)\right)}.
	 \end{displaymath}
	 \noindent where $\mathbf{a}$ is learned, the superscript $t$ represents transposition and $\|$ is the concatenation operation.
	 
	 \subsubsection{GATv2}
	 The GATv2 is a modification of GAT, with the attention coefficients \(\alpha_{i,j}\) computed as follows
	 \begin{displaymath}
	   \alpha_{i,j} = \frac{ \exp\left(\mathbf{a}^{t}\mathrm{LeakyReLU}\left( [\mathbf{\Theta}\mathbf{x}_i \, \| \, \mathbf{\Theta}\mathbf{x}_j] \right)\right)} {\sum_{k \in \mathcal{N}(i) \cup \{ i \}} \exp\left(\mathbf{a}^{t}\mathrm{LeakyReLU}\left( [\mathbf{\Theta}\mathbf{x}_i \, \| \, \mathbf{\Theta}\mathbf{x}_k] \right)\right)}
	 \end{displaymath}	 
	 GATv2 is a dynamic graph attention variant that is strictly more expressive than the GAT which is a static graph attention variant \cite{brody2022how}. 
	 
	 \subsubsection{GraphSAGE}
	 The GraphSAGE (SAmple and aggreGatE) learns a function that generates embeddings by sampling and aggregating features from local neighbors, which can efficiently generate node representation for previously unseen data \cite{Hamilton2017}. 
	 The core layer of GraphSAGE is as follows:
	 \begin{displaymath}
	   \mathbf{x}_{i}^{(k)} = \sigma \left( W^{(k)} \cdot \left(\mathbf{x}_{i}^{(k-1)} \| \text{AGGREGATE} (\{\mathbf{x}_{j}^{(k-1)} | j \in \mathcal{N}_{i} \}  ) \right)  \right) 
	 \end{displaymath}
	 \noindent where $W^{(k)}$ is the weight matrix of $k$-th layer, and the AGGREGATE method could be mean aggregator, LSTM aggregator, or pooling aggregator \cite{Hamilton2017}. 
	 

	 \subsubsection{Graph transformer}
	 The graph transformer networks are the integration of graph networks and transformer \cite{Vaswani2017}. The graph transformer (GraphTF) layer identifies useful connections, learns a soft selection, and composites an effective node representation \cite{ijcai2021p214}. 
	 \begin{displaymath}
	   \mathbf{x}_{i}^{(k)} = 
	   W^{(k)}_{1} \mathbf{x}_{i}^{(k-1)} + \sum_{j \in \mathcal{N}(i)} \alpha_{i, j} W^{(k)}_{2} \mathbf{x}_{j}^{(k-1)}
	 \end{displaymath}
	 \noindent where the attention coefficients $\alpha_{i, j}$ are computed via multi-head dot product attention \cite{ijcai2021p214,Vaswani2017}.
	 
	 %
	 %
	 
	 \subsubsection{Topology adaptive graph convolutional network}
	 The topology adaptive graph convolutional network (TAGCN) is a generalization of GCN and adopts a set of fixed-size learnable filters to perform convolutions on graphs, which is adaptive to the topology of the graph \cite{du2018topology}.
	 The core layer of TAGCN is as follows
	 \begin{displaymath}
	   X^{(k)}=\sum_{p=1}^{P}\left(\tilde{D}^{-\frac{1}{2}} \tilde{A} \tilde{D}^{-\frac{1}{2}}\right)^{p} X^{(k-1)} \mathbf{\Theta}_{p}^{(k-1)}
	 \end{displaymath} 
	 where the $p$ denotes the number of hops on the graph.
	 %
	 
	 \section{Experiments}\label{sec:exp}
	 
	 \subsection{Results on SRE14 i-vector database}
	 The SRE14 i-vector challenge \cite{Greenberg2014} takes vectors instead of speech as input to compare different speaker verification backends fairly. The dataset is gender independent and contains 1306 speaker models, 9634 test segments, and 12582004 trials. 
	 Each speaker model has 5 i-vectors. 
	 Trials are randomly divided into a progress subset (40$\%$) and an evaluation subset (60$\%$). 
	 In addition, NIST provided a development set containing 36572 i-vectors. 
	 All i-vectors are 600 dimensional. 
     We take both training and test datasets to construct graphs with nearest neighbor algorithms. 
	 The development data with labels are used to train LDA, PLDA, and GNN backends.
	 The default dimension of LDA and PLDA are 250 and 50, respectively. 
	 The GNN layer is implemented by the Pytorch geometric.
	 Unless otherwise specified, our model architecture includes two layers of GNN+BN, a linear layer, and a softmax output followed by a cross-entropy loss. 
	 The g-vectors (See Fig \ref{fig:gvec}) are extracted for the final cosine decision.
	 The model is trained with $600$ epochs, a fixed learning rate of $10^{-4}$, and a weight decay parameter of $5\times10^{-4}$.
	 
	 The TABLE \ref{tab:cmp-bckend-sre14} shows the comparison results of cosine, LDA, LDA+PLDA, DBL, and our proposed GNN backends on the SRE14 dataset.
	 From the table, we can see that the EER and $\text{minDCF}_{14}$ of our proposed GNN backend are $1.69\%$ and $0.238$ on the Progress set, and $1.55\%$ and $0.218$ on the Evaluation set, which gains $22.4\%$, $0.4\%$, $25.8\%$, and $4.3\%$ relative performance improvement compared with the LDA+PLDA, respectively.
	 
	 \begin{table}[!htp]
	   \caption{Comparison of different backends on SRE14}
	   \label{tab:cmp-bckend-sre14}
	   \centering
	   \begin{tabular}{lcccc}
	     \toprule
	     \multirow{2}{*}{Backend} 
	     & \multicolumn{2}{c}{Progress Set} & \multicolumn{2}{c}{Evaluation Set} \\ 
	     \cmidrule(r){2-3} \cmidrule(r){4-5}
	     & EER[$\%$] & $\text{minDCF}_{14}$ & EER[$\%$] & $\text{minDCF}_{14}$ \\ \hline
	     Cosine \cite{Greenberg2014} & 4.78 & 0.386 & 4.46 & 0.378 \\ \hline
	     LDA \cite{Greenberg2014}    & 4.44 & 0.343 & 4.04 & 0.332 \\ \hline
	     LDA+PLDA \cite{GarciaRomero2011}    & 2.18 & 0.239 & 2.09 & 0.228 \\ \hline
	     DNN-3L \cite{Ghahabi2017}       & 4.55 & 0.305 & 4.11 & 0.300 \\ \hline
	     GNN-backend(GAT) & \bf{1.69} & \bf{0.238} & \bf{1.55} & \bf{0.218} \\    
	     \bottomrule
	   \end{tabular}
	 \end{table}
	 
	 \subsection{Ablation study on SRE14 i-vector database}
	 From the TABLE \ref{tab:cmp-var-sre14}, we find that the GAT has the best results in EER and the GCN has the best results in $\text{minDCF}_{14}$.
	 The GAT learns adaptive weights to edges through the attention mechanism, which means it can filter more effective neighbors for auxiliary decisions.
	 The GCN is good at capturing global information on the graph, and we think it's the reason for its good performance.
	 The structure of GATv2 is similar to the GAT, and its performance is also approximate to the GAT.
	 Both the GCN and GraphSAGE (mean aggregator in our experiment) have a similar aggregation way. 
	 However, the GCN takes advantage of the adjacency matrix to normalize the node and its neighbors, which may learn the more robust potential pattern.
	 Similar to the GAT, for a node on the graph, the GraphTF also learns adaptive weights from its neighbors. 
	 Introducing multi-heads in the GraphTF brings more freedom to fit the local structure information, which may be adverse to obtaining a model with good generalization ability, especially under limited training data.
	 The TGGCN considers multi-hops (3 hops in our experiments), which is suitable for the social network or recommendation system. 
	 But it also brings instability, and we guess that's the reason for the poor results of TAGCN. 
	 According to the above analysis, we do the following ablation experiments using the GAT.
	 
	 \begin{table}[!htp]
	   \caption{Comparison of variants of GNN on SRE14}
	   \label{tab:cmp-var-sre14}
	   \centering
	   \begin{tabular}{lcccc}
	     \toprule
	     \multirow{2}{*}{Variant} & \multicolumn{2}{c}{Progress Set} & \multicolumn{2}{c}{Evaluation Set} \\
	     \cmidrule(r){2-3} \cmidrule(r){4-5}
	     & EER[$\%$] & $\text{minDCF}_{14}$ & EER[$\%$] & $\text{minDCF}_{14}$ \\ \hline
	     GCN\cite{kipf2017semisupervised}                 &1.85 & \bf{0.227} & 1.78 & \bf{0.208} \\ \hline
	     GAT\cite{velickovic2017graph} & \bf{1.69} & 0.238 & \bf{1.55} & 0.218 \\ \hline
	     GATv2\cite{brody2022how}         & 1.91 & 0.228 & 1.71 & 0.212 \\ \hline
	     GraphSAGE\cite{Hamilton2017}   & 2.99 & 0.272 & 2.73 & 0.252 \\ \hline
	     GraphTF\cite{ijcai2021p214}   & 2.92 & 0.253 & 2.66 & 0.248 \\ \hline
	     TAGCN\cite{du2018topology}           & 2.78 & 0.230 & 2.62 & 0.221 \\ 
	     \bottomrule
	   \end{tabular}
	 \end{table}

  We study the construction method of graphs, e.g., nodes and edges, subsequently.
  The TABLE \ref{tab:cmp-con-sre14} shows that the performance is best when the nodes are 250-dimensional vectors after LDA reduction and edges are built by the PLDA (50 dimensions). 
  We conclude that the dimension deduction is necessary for graph building.
 
 \begin{table}[!htp]
 	\caption{Comparison of different graph construction on SRE14}
 	\label{tab:cmp-con-sre14}
 	\centering
 	\begin{tabular}{cllcc}
 		\toprule
 		Dataset & Node & Edge & EER[$\%$] & $\text{minDCF}_{14}$ \\ \hline
 		\multirow{4}{*}{Progress Set} 
 		& 600      & Cosine(600) & 5.14 & 0.449 \\ \cline{2-5}
 		& 600      & Cosine(LDA 250) & 3.46 & 0.371 \\ \cline{2-5}
 		& 250(LDA) & Cosine(600) & 2.79 & 0.264 \\ \cline{2-5}
 		& 250(LDA) & Cosine(LDA 250) & 2.05 & 0.275 \\ \cline{2-5}
 		& 250(LDA) & PLDA(50) &  \bf{1.69} & \bf{0.238} \\ \hline
 		\multirow{4}{*}{Evalution Set} 
 		& 600      & Cosine(600) & 4.74 & 0.449 \\ \cline{2-5}
 		& 600      & Cosine(LDA 250) & 3.18 & 0.371 \\ \cline{2-5}
 		& 250(LDA) & Cosine(600) & 2.66 & 0.257 \\ \cline{2-5}
 		& 250(LDA) & Cosine(LDA 250) & 1.81 & 0.263 \\ \cline{2-5}
 		& 250(LDA) & PLDA(50) & \bf{1.55} & \bf{0.218} \\ 
 		\bottomrule
 	\end{tabular}
 \end{table}

	 
	 From the TABLE \ref{tab:cmp-thres-sre14}, we find that when the threshold is bewteen 4 and 10, the proposed method will maintain relatively stable and good performance.
	 If the threshold is too low, there will be too many neighbors for a node, which makes the local graph structure tend to be consistent, and this is harmful to the classification task. 
	 If the threshold is too high, there will be too few neighbors for a node, which makes the training data too sparse. 
	
		 \begin{table}[!htp]
		\caption{GAT backend with different thresholds on SRE14}
		\label{tab:cmp-thres-sre14}
		\centering
		\begin{tabular}{lcccc}
			\toprule
			\multirow{2}{*}{Threshold} & \multicolumn{2}{c}{Progress Set} & \multicolumn{2}{c}{Evaluation Set} \\
			\cmidrule(r){2-3} \cmidrule(r){4-5}
			& EER[$\%$] & $\text{minDCF}_{14}$ & EER[$\%$] & $\text{minDCF}_{14}$ \\ \hline
			2          & 1.81 & 0.354 & 1.68 & 0.336 \\ \hline
			4          & \bf{1.65} & 0.302 & \bf{1.50} & 0.282 \\ \hline
			6          & \bf{1.65} & 0.261 & 1.52 & 0.244 \\ \hline
			8          & 1.69 & 0.236 & 1.55 & 0.218 \\ \hline
			10         & 1.81 & \bf{0.226} & 1.71 & \bf{0.209} \\ \hline
			12         & 1.95 & 0.230 & 1.79 & 0.214 \\ 
			\bottomrule
		\end{tabular}
	\end{table}
 

	 From the TABLE \ref{tab:cmp-layers-sre14}, we learn that when the layer is $2$, the performance is the best. 
	 When the layer is $1$, the shallow model cannot effectively learn the nonlinear structure on the graph. 
	 And, it exists over-smoothing if too many layers are adopted. 
	 
	 \begin{table}[!htp]
	   \caption{GAT backend with different layers on SRE14}
	   \label{tab:cmp-layers-sre14}
	   \centering
	   \begin{tabular}{lcccc}
	     \toprule
	     \multirow{2}{*}{Layer} & \multicolumn{2}{c}{Progress Set} & \multicolumn{2}{c}{Evaluation Set} \\
	     \cmidrule(r){2-3} \cmidrule(r){4-5}
	     & EER[$\%$] & $\text{minDCF}_{14}$ & EER[$\%$] & $\text{minDCF}_{14}$ \\ \hline
	     1           & 1.70 & 0.245 & 1.62 & 0.227 \\ \hline
	     2           & \bf{1.69} & \bf{0.236} & \bf{1.55} & \bf{0.218} \\ \hline
	     3           & 2.16 & 0.249 & 1.92 & 0.233 \\ \hline
	     4           & 2.22 & 0.258 & 2.15 & 0.241 \\
	     \bottomrule
	   \end{tabular}
	 \end{table}
 
	 
	 \subsection{Results on VoxCeleb1-O, VoxCeleb1-E, and VoxCeleb1-H}

Our proposed method is also evaluated on the VoxCeleb1 dataset, using the development set of VoxCeleb2 \cite{chung2018voxceleb2} as the training data for our frontend x-vector extractor. 
The frontend model utilizes a TDNN network, a statistic pooling layer, and a fully-connected layer, optimized with the AAM-Softmax loss \cite{Deng2019} function. 
The evaluation is performed on all three official trial lists: VoxCeleb1-O, VoxCeleb1-E, and VoxCeleb1-H\cite{vox2019}. The features are extracted using 80-dimensional Fbank features with voice activity detection, augmented with MUSAN\cite{Snyder2015} and RIRs noise sources, and trained with an Adam optimizer with a $10^{-3}$ initial learning rate and a weight decay of $10^{-4}$. 
The parameters of GCN backends are re-tuned in a similar way, as mentioned earlier. 
The experiment results shown in TABLE \ref{tab:cmp-var-vox} demonstrate the effectiveness of graph neural network (GNN) variants on the VoxCeleb1 dataset. 
Specifically, GCN outperforms all other GNN-based methods with the lowest EER and $\text{minDCF}_{0.01}$. 
These results suggest that GNN-based methods can effectively extract speaker-related discriminant information and are promising for speaker verification tasks.
	 

\begin{table}[!htp]
\centering
\caption{Comparison results GNN on VoxCeleb1 (EER[$\%$] and $\text{minDCF}_{0.01}$)}
\label{tab:cmp-var-vox}
\begin{tabular}{lllllll}
\toprule
\multicolumn{1}{c}{\multirow{2}{*}{Variant}} & \multicolumn{2}{l}{VoxCleb1-O} & \multicolumn{2}{l}{VoxCeleb1-E} & \multicolumn{2}{l}{VoxCeleb-H} \\ \cline{2-7} 
\multicolumn{1}{c}{}                         & EER           & $\text{minDCF}$         & EER           & $\text{minDCF}$          & EER           & $\text{minDCF}$         \\ \hline
Cosine                                       & 2.70          & 0.302          & 2.58          & 0.291           & 4.44          & 0.410          \\ \hline
LDA-PLDA                                     & 3.13          & 0.382          & 3.29          & 0.391           & 6.00          & 0.568          \\ \hline
GCN                                          & \bf{0.46}          & \bf{0.054}          & \bf{0.68}          & \bf{0.090}           & \bf{1.12}          & \bf{0.127}          \\ \hline
GAT                                          & 1.83          & 0.080          & 0.97          & 0.198           & 1.70          & 0.357          \\ \hline
GATv2                                        & 1.93          & 0.086          & 1.35          & 0.200           & 2.12          & 0.334          \\ \hline
GraphSAGE                                    & 1.27          & 0.083          & 1.01          & 0.201           & 1.78          & 0.374          \\ \hline
GraphTF                                      & 1.56          & 0.103          & 1.35          & 0.219           & 2.15          & 0.405          \\ \hline
TAGCN                                        & 1.08          & 0.076          & 0.97          & 0.187           & 1.69          & 0.347          \\ 
\bottomrule
\end{tabular}
\end{table}

	 
	 \section{Conclusion}
	 We propose a graph neural network (GNN) backend for speaker recognition. 
	 The proposed method can capture the structural relation among extracted i-vectors or x-vectors on a graph and thus allows us to take advantage of more information for classification compared with analyzing them in isolation.
	 The embeddings extracted from the GNN, named g-vectors, are excellent representations and preserve rich graph properties in a low-dimensional Euclidean space, which contains more discriminant information.
	 The detailed experimental results on the SRE14 i-vector and VoxCeleb1-O, VoxCeleb1-E, and VoxCeleb1-H datasets demonstrate that our proposed GNN backend is very effective.
	 
	 \newpage
	 \bibliography{IEEEabrv,sre_graph_20230408}
  \bibliographystyle{IEEEtran}
\end{document}